\documentclass[a4paper,12pt]{article}
\usepackage{jcappub}
\usepackage[symbol]{footmisc}
\usepackage{amsmath,amssymb}
\usepackage{revsymb}
\usepackage{orcidlink}
\usepackage{bm}
\newcommand{\nc}{\newcommand*}

\nc{\al}{\alpha}
\nc{\s}{\sigma}
\nc{\kp}{\kappa}
\nc{\dt}{\delta}
\nc{\Dt}{\Delta}
\nc{\Ld}{\Lambda}
\nc{\p}{\partial}
\nc{\Gm}{\Gamma}
\nc{\om}{\omega}
\nc{\Om}{\Omega}
\nc{\rd}{\mathrm{d}}

\def\({\left(}
\def\){\right)}
\def\[{\left[}
\def\]{\right]}
\def\e{\begin{equation}}
\def\q{\end{equation}}
\def\be{\begin{equation}}
\def\ee{\end{equation}}
\def\m{\begin{eqnarray}}
\def\n{\end{eqnarray}}
\def\beq{\begin{eqnarray}}
\def\eeq{\end{eqnarray}}
\nc{\Eq}[1]{Eq.~\eqref{#1}}     
\nc{\Fig}[1]{Fig.~\ref{#1}}     
\nc{\Table}[1]{Table~\ref{#1}}  
\nc{\Sec}[1]{Sec.~\ref{#1}}
\nc{\Msun}{M_\odot}             
\nc{\Ogw}{\Omega_{\mathrm{GW}}}
\nc{\gpcyr}{\mathrm{Gpc}^{-3}\,\mathrm{yr}^{-1}}
\nc{\lvc}{LIGO/Virgo} 
\nc{\SNR}{\mathrm{SNR}} 
\nc{\rhoGW}{\rho_{\mathrm{GW}}}
\nc{\vd}{\vec{d}}
\nc{\av}[1]{\langle #1 \rangle} 
\nc{\km}{\mathrm{km}}
\nc{\Mpc}{\mathrm{Mpc}}
\nc{\Tobs}{T_{\mathrm{obs}}}
\nc{\fyr}{f_{\mathrm{yr}}}
\nc{\mH}{\mathcal{H}}
\nc{\cosuz}{\cos\frac{ux}{\sqrt{3}}}
\nc{\sinuz}{\sin\frac{ux}{\sqrt{3}}}
\nc{\cosvz}{\cos\frac{vx}{\sqrt{3}}}
\nc{\sinvz}{\sin\frac{vx}{\sqrt{3}}}
\nc{\addref}{[\textcolor{red}{add ref}] } 
\nc{\eg}{\textit{e.g.~}}
\nc{\app}{\approx}
\nc{\hf}{\frac{1}{2}}
\nc{\discuss}{\textcolor{red}{Add discussion here!}}
\nc{\red}[1]{\textcolor{red}{#1}}
\allowdisplaybreaks[3]

\begin{document}

\title{On the Gauge Invariance of Secondary Gravitational Waves}

\author[a]{Chen Yuan,\orcidlink{0000-0001-8560-5487}}
\author[b,c,*]{Yizhou Lu,\note{Corresponding author.}\orcidlink{0000-0002-1150-244X}}
\author[d,e,*]{Zu-Cheng Chen\orcidlink{0000-0001-7016-9934}}
\author[f,*]{Lang~Liu\orcidlink{0000-0002-0297-9633}}

\affiliation[a]{CENTRA, Departamento de Física, Instituto Superior Técnico – IST, Universidade de Lisboa – UL, Avenida Rovisco Pais 1, 1049–001 Lisboa, Portugal}
\affiliation[b]{Shanghai Institute for Mathematics and Interdisciplinary Sciences, Shanghai 200000, China}
\affiliation[c]{School of Mathematical Sciences, Fudan University, Shanghai 200000, China}
\affiliation[d]{Department of Physics and Synergetic Innovation Center for Quantum Effects and Applications, Hunan Normal University, Changsha, Hunan 410081, China}
\affiliation[e]{Institute of Interdisciplinary Studies, Hunan Normal University, Changsha, Hunan 410081, China}
\affiliation[f]{Faculty of Arts and Sciences, Beijing Normal University, Zhuhai 519087, China}

\emailAdd{chenyuan@tecnico.ulisboa.pt}
\emailAdd{luyz@simis.cn}
\emailAdd{zuchengchen@hunnu.edu.cn}
\emailAdd{liulang@bnu.edu.cn}

\abstract{
Second-order tensor perturbations induced by primordial fluctuations play a crucial role in probing small-scale physics, but gauge dependence of their energy density has remained a fundamental challenge in cosmological perturbation theory. We address this issue by introducing a boundary condition-based filtering method that extracts physical radiation through the Sommerfeld criterion. We demonstrate that after filtering non-physical modes, the energy density of secondary gravitational waves becomes gauge-invariant and exhibits physically consistent behavior in the sub-horizon limit. This approach provides a unified framework for both adiabatic and isocurvature perturbations, enhancing theoretical predictions and observational signatures of early universe physics.
}
	
\maketitle
\section{Introduction}
Gravitational waves (GWs), one of the most profound predictions of general relativity, have revolutionized astrophysics and cosmology since the detection of binary black hole merger events by the LIGO-Virgo collaboration~\cite{LIGOScientific:2016dsl,LIGOScientific:2018mvr,LIGOScientific:2020ibl,LIGOScientific:2021usb,KAGRA:2021vkt}. 
Beyond direct detection, GWs provide essential insights into early universe physics through both linear and higher-order tensor perturbations, particularly through scalar-induced gravitational waves (SIGWs)~\cite{Saito:2008jc,Sasaki:2018dmp,Chen:2019xse,Vaskonen:2020lbd,DeLuca:2020agl,Yuan:2021qgz,Domenech:2021ztg,Domenech:2023jve}.

While the first-order transverse-traceless tensor perturbations are invariant under infinitesimal diffeomorphisms (gauge-invariant) in a cosmological background, the general covariance of Einstein equation dictates the gauge dependence of higher-order perturbations~\cite{Matarrese:1997ay,Noh:2003yg}.
This fundamental issue has prompted various attempts to define gauge-invariant quantities, similar to approaches used in Schwarzschild spacetime~\cite{Spiers:2023mor}.
However, the cosmological scenario presents unique challenges due to the intrinsic coupling between scalar and tensor perturbations in the entire spacetime, making it particularly difficult to define physical observables such as the dimensionless energy density $\Omega_{\mathrm{GW}}$. Direct calculations reveal that $\Omega_{\mathrm{GW}}$ for SIGWs exhibits gauge dependence and even diverges in certain gauges~\cite{Gong:2019mui,Tomikawa:2019tvi,DeLuca:2019ufz,Inomata:2019yww,Yuan:2019fwv,Lu:2020diy,Ali:2020sfw,Yuan:2024qfz}.

The gauge discrepancies in SIGWs were first identified numerically for adiabatic perturbations~\cite{Hwang:2017oxa} and later confirmed through analytical calculations in various gauges~\cite{Gong:2019mui,Tomikawa:2019tvi,DeLuca:2019ufz,Inomata:2019yww,Yuan:2019fwv,Lu:2020diy,Ali:2020sfw}. Recent studies have shown that similar gauge dependence affects SIGWs generated by isocurvature perturbations~\cite{Yuan:2024qfz}. Previous attempts to address this issue include arguments about gauge suitability based on source term activity~\cite{Domenech:2020xin} and the introduction of canonical observer tetrads using Newman-Penrose formalism~\cite{Cai:2021jbi,Newman:1961qr}. However, a complete resolution with direct physical interpretation remains elusive.

The gauge dependence of $\Omega_{\mathrm{GW}}$ presents a fundamental challenge for cosmological observations. Since $\Omega_{\mathrm{GW}}$ impacts cosmic evolution, gauge-dependent values would lead to observer-dependent cosmic histories, an unphysical scenario. Moreover, the divergence of $\Omega_{\mathrm{GW}}$ in certain gauges implies unphysical GW domination. Resolving these issues is crucial for establishing a robust theoretical framework for higher-order GWs and ensuring reliable observational predictions.

In this paper, we demonstrate that the observed gauge dependence stems from unphysical modes that violate the Sommerfeld boundary conditions. We introduce a boundary condition-based filtering method to isolate physical radiation, showing that the filtered $\Omega_{\mathrm{GW}}$ is gauge-invariant. We apply this method to both adiabatic and isocurvature induced SIGWs in an arbitrary gauge, establishing a unified framework for physical gravitational radiation. Throughout this paper, we use natural units with $G=c=1$.

\section{SIGWs and Gauge Dependence}
We begin with the second-order Einstein equation in Newtonian gauge.
Other gauges are acquired by performing a gauge transformation thereafter (see Appendix for details of metric perturbations up to second order). Our master equation is
\begin{equation}\label{eomhij}
(a h_{ij})'' - \left( \nabla^2 + \frac{a''}{a} \right) (a h_{ij}) = 4 a \mathcal{T}_{ij}^{mn}S_{mn},
\end{equation}
where $h_{ij}$ denotes the second-order transverse-traceless tensor mode, $a$ is the scale factor, and $\mathcal{T}_{ij}^{mn}$ is the transverse-traceless projection operator. Here, a prime represents the derivative with respect to conformal time. 
We consider a general cosmological background with equation of state $w$ and sound speed $c_s$.
Using the Green's function method, this equation can be solved in Fourier space as 
\begin{equation}\label{hsol}
h_k^{\lambda}(\eta) = \frac{4}{k^2 } \int \frac{\mathrm{d} ^3 \mathbf{p}}{(2\pi)^{3}} 
\mathbf{e}_{ij}^{\lambda}(\mathbf{k}) p_i p_j X_\mathbf{p}X_{\mathbf{k} - \mathbf{p}} I_X^N(u, v, x),
\end{equation}
where $\mathbf{e}_{ij}^{\lambda}(\mathbf{k})$ is the GW polarization tensor ($\lambda=\{+,\times\}$), normalized such that $\mathbf{e}_{ij}^{\lambda}\mathbf{e}_{ij}^{\lambda'}=\delta^{\lambda\lambda'}$. 
The kernel function $I_X$ encodes the time evolution of the source through the Green's function integral and has known analytical forms~\cite{Kohri:2018awv}.
For convenience, we introduce the dimensionless variables $x\equiv k\eta$, $u=p/k$, and $v=|\mathbf{k}-\mathbf{p}|/k$. 
The perturbation type $X$ represents either primordial comoving curvature perturbations ($X=\zeta$, adiabatic case) or primordial gauge-invariant entropy perturbations ($X=S$, isocurvature case).
The GW energy density can be expressed as
\begin{equation}\label{OgwN}
\begin{aligned}
    &\Omega^{N}_{\mathrm{GW}}(k) = \frac{(1+3w)^2}{24} \int_0^\infty \mathrm{d}u \int_{|1-u|}^{1+u} \mathrm{d}v \frac{v^2}{u^2} 
    \mathcal{P}_X(uk)\mathcal{P}_X(vk)\\
    &\quad\times\left[1 - \left( \frac{1+v^2-u^2}{2v} \right)^2 \right]^2x^2\overline{(I_X^N(u,v,x))^2},~(x\to\infty). 
\end{aligned}
\end{equation}
Here, for the adiabatic case, $\mathcal{P}_{X}=\mathcal{P}_{\zeta}$ is the primordial power spectrum of curvature perturbations and $I_X^N = I_{\mathrm{AD}}$; for the isocurvature case, $\mathcal{P}_{X}=\mathcal{P}_{S}$ is the primordial power spectrum of entropy perturbations and $I_X^N = I_{\mathrm{ISO}}$. An overline on the kernel function denotes oscillating average such that $\overline{\sin^2 x}=\overline{\cos^2x}=1/2$. Note that Eq.~(\ref{OgwN}) is valid for both adiabatic case and isocurvature case and taking the sub-horizon limit, $x\to\infty$ is equivalent to evaluating $\Omega_{\mathrm{GW}}$ at the equality time.

The kernel function can be decomposed into three components:
\begin{equation}\label{Iuvsplit}
I_{\mathrm{X}}^N=I_{\mathrm{X,Y}}^N(u,v,x) +I_{\mathrm{X,J}}^N(u,v,x)+\Delta I_{\mathrm{X}}^N(u,v,x).
\end{equation}
For both adiabatic and isocurvautre perturbations, the first two components take the form
\begin{equation}\label{IuvNewton}
\begin{aligned}
I_{\mathrm{AD,Y}}^N&=x^{-\beta}
\mathcal{T}_{X,Y}(u,v,w)Y_{\beta}(x),\\
I_{\mathrm{AD,J}}^N&=x^{-\beta}
\mathcal{T}_{X,J}(u,v,w)J_{\beta}(x),
\end{aligned}
\end{equation}
where $\beta=3(1-w)/[2(1+3w)]$, and $Y_{n}(x)$ and $J_n(x)$ are the Bessel functions of the first and second kind respectively. 
The transfer functions $\mathcal{T}_{X,Y/J}$ have lengthy expressions available in \cite{Domenech:2019quo} for adiabatic perturbations and in \cite{Domenech:2024wao} for isocurvature perturbations.
In the sub-horizon limit ($x\gg1$), the Bessel functions asymptotically behave as
\begin{equation}
\left\{
\begin{array}{l}
J_n(x\gg1) \\
Y_n(x\gg1)
\end{array}
\right\}
\simeq \sqrt{\frac{2}{\pi x}}
\left\{
\begin{array}{l}
\cos \\
\sin
\end{array}
\right\}
\left(x-\frac{n\pi}{2}-\frac{\pi}{4}\right),
\end{equation}
so that $I_{\mathrm{X,Y}}^N$ and $I_{\mathrm{X,J}}^N$ describe GWs oscillating as $x^{-\beta-1/2}\sin x$ or $x^{-\beta-1/2}\cos x$ respectively.

The last component in \Eq{Iuvsplit} lacks a general analytical expression for arbitrary $w$ in the adiabatic case. 
However, during radiation domination ($w=1/3$), it takes the explicit form of 
\begin{equation}\label{IuvADN}
    \begin{aligned}
    &\Delta I_{\mathrm{AD}}^N(u,v,x)=\frac{9}{u^3 v^3 x^4} \bigg[ 
      6 \sin\frac{u x}{\sqrt{3}} \sin\frac{v x}{\sqrt{3}} \\
     &+ x^2 \left( 
        2 u v \cos\frac{u x}{\sqrt{3}} \cos\frac{v x}{\sqrt{3}} + 
        (-3 + u^2 + v^2) \sin\frac{u x}{\sqrt{3}} \sin\frac{v x}{\sqrt{3}} 
    \right)\\
    &  
    - 2 \sqrt{3} x \big( 
        v \cos\frac{v x}{\sqrt{3}} \sin\frac{u x}{\sqrt{3}} + 
        u \cos\frac{u x}{\sqrt{3}} \sin\frac{v x}{\sqrt{3}} 
    \big) 
\bigg].
\end{aligned}
\end{equation}
For isocurvature perturbation, a general expression for $\Delta I_{\mathrm{ISO}}^N$ is also absent in an arbitrary cosmological background.
While an explicit form exists during radiation domination, it is algebraically complex and not necessary for our analysis. Importantly, both components vanishes in the sub-horizon limit in a genearl background~\cite{Domenech:2021and,Domenech:2023jve,Yuan:2024qfz} 
\begin{equation}
    \Delta I_{\mathrm{AD}}^N(u,v,x\to\infty )=0,~\Delta I_{\mathrm{ISO}}^N(u,v,x\to\infty )=0.
\end{equation}

Now we consider an arbitrary first-order gauge transformation, $x^\mu \to \tilde{x}^\mu =x^\mu +\xi^\mu$, where $\xi^\mu=(\alpha,\partial^{i}L)$. 
Such transformations modify the kernel function as~\cite{Lu:2020diy}
\begin{equation}
    I_X(u,v,x) 
    \to I_X(u,v,x) +I_\chi(u,v,x),
\end{equation}
where, for a general background, $I_\chi(u,v,x)$ takes the form
\begin{equation}\label{Ichi}
\begin{aligned}
    &I_{\chi} = -\frac{1}{4uv}\Bigg[2 T_\alpha(ux)T_\alpha(vx) + \frac{1 - u^2 - v^2}{uv} T_L(ux)T_L(vx) 
      \\
    &-4\left( \frac{u}{v} T_N(ux)T_L(vx) + \frac{v}{u} T_N(vx)T_L(ux) \right)\\
    &+ \frac{8}{x(1+3w)}\left( \frac{1}{v} T_\alpha(ux)T_L(vx) + \frac{1}{u} T_L(ux)T_\alpha(vx) \right) 
\Bigg].
\end{aligned}
\end{equation}
Here, $T_{N}$ represents the transfer function of $\phi$ in Newtonian gauge (see Appendix for details), while $T_\alpha$ and $T_{L}$ are the transfer functions of $\alpha$ and $L$ respectively.
This transformation is valid for both the adiabatic and isocurvature perturbations,
allowing the derivation of kernel functions in arbitrary gauges through appropriate choices of transformation parameters.

A fundamental challenge emerges from this gauge freedom: the GW energy density $\Omega_{\rm GW}$ exhibits explicit gauge dependence~\cite{Gong:2019mui,Tomikawa:2019tvi,DeLuca:2019ufz,Inomata:2019yww,Yuan:2019fwv,Lu:2020diy,Ali:2020sfw,Yuan:2024qfz}. To resolve this physical inconsistency, we introduce a boundary condition-based filtering method in the following section.

\section{Boundary Condition-Based Filtering and Canonical Observer}\label{method}
The Sommerfeld boundary condition is a fundamental criterion for physical waves, requiring that they exhibit no incoming flux at infinity. Mathematically, the GWs around flat spacetime must satisfy the following asymptotic behavior~\cite{Maggiore:2007ulw}:
\begin{equation}\label{bc}
    \lim_{t\to\infty} (\partial_\eta+\partial_r)(r a h_{ij})=0.
\end{equation} 
The Sommerfeld boundary condition suggests that GWs, far from the source, behave as outgoing spherical waves (in our notation, this corresponds to $h_{ij}\sim x^{-\beta-1/2}$ in the sub-horizon limit) propagating at the speed of light. Modes that violate this condition, such as those with divergent behavior, do not represent physical radiation.
The basic idea of our proposed boundary condition-based filtering method is that the tensor perturbations  $h_{ij}$ cannot be simply regarded as GWs. A tensor perturbation consists of both non-physical waves and GWs, and only modes that satisfy the Sommerfeld boundary condition can be physically taken as GWs by a canonical observer. 
{Such an idea of decomposition the second-order tensor modes into the propagating GW part (proportional to $\sin(k\eta)$ or $\cos(k\eta)$) and the non-GW part (proportional to $\sin(kx/\sqrt{3})$ and $\sin(kx/\sqrt{3})$), was first pointed out in \cite{Inomata:2019yww} for SIGWs.} This is also argued in~\cite{Cai:2021jbi} where the authors pointed out that only $h_{ij}$ in the canonical 3-metric can be physically interpreted as GWs.


To compute $\Omega_{\mathrm{GW}}$, one must identify modes in $h_{ij}$ that both decay as spherical waves and propagate at the speed of light, a criterion that reduces to analyzing the kernel function's asymptotic behavior in the sub-horizon limit [see Eq.~(\ref{hsol})]. Previous calculations of $\Omega_{\mathrm{GW}}$ for SIGWs~\cite{Gong:2019mui,Tomikawa:2019tvi,DeLuca:2019ufz,Inomata:2019yww,Yuan:2019fwv,Lu:2020diy,Ali:2020sfw,Yuan:2024qfz}, which directly interpreted the second-order tensor mode $h_{ij}$ as GWs, led to gauge-dependent divergences ($x^6$ in uniform density gauge and $x^2$ in total matter gauge). We demonstrate that applying boundary condition-based filtering to identify physical modes in $h_{ij}$ yields a gauge-invariant $\Omega_{\mathrm{GW}}$ for SIGWs, eliminating all divergences.

Firstly, we apply the boundary condition-based filtering method to Newtonian gauge. In the previous section, we show that $I_{\mathrm{X,Y/J}}^N$ oscillates and decays as $x^{-\beta-1/2}$. This is consistent with the time evolution of tensor perturbations in the absence of source terms. On the other hand, $I_{\mathrm{X,Y/J}}^N$ oscillates as $\sin x$ or $\cos x$, indicating these modes are light-speed-propagating modes. Therefore, $I_{\mathrm{X,Y/J}}^N$ represents physical waves that propagate at the speed of light and decay as spherical waves in $h_{ij}$ and hence, can be physically regarded as GWs by a canonical observer.
However, we see that, for adiabatic perturbations during RD, $\Delta I_{\mathrm{AD}}^N$ has $1/x^2$, $1/x^3$ and $1/x^4$ dependence and it contains oscillating terms such as $\sin(ux/\sqrt{3})$ and $\cos(vx/\sqrt{3})$, indicating that these modes do not propagate at the speed of light. 
From the perspective of quantum field theory, $ux$ and $vx$ represent the virtual momenta $p$ and $|\mathbf{k}-\mathbf{p}|$ respectively, while $1/\sqrt{3}$ indicates the propagation at the sound speed during RD. The time dependence of such oscillating terms in $\Delta I_{\mathrm{AD}}^N$ reflects the propagation of virtual modes, rather than physical waves.
Hence, $\Delta I_{\mathrm{AD}}^N$ represents pure non-physical radiation that cannot be treated as GWs. This non-physical radiation in the second-order tensor modes arises due to gauge choice. However, the non-physical radiation, both $\Delta I_{\mathrm{AD}}^N$ and $\Delta I_{\mathrm{ISO}}^N$, vanishes in the sub-horizon limit and hence does not affect $\Omega_{\mathrm{GW}}$ observed by a canonical observer.
This result in the Newtonian gauge is also valid for a general cosmological background as shown in \cite{Domenech:2019quo} for adiabatic perturbations and in \cite{Domenech:2024wao} for isocurvature perturbations.

Secondly, notice that the extra kernel function $I_\chi(u,v,x)$ in an arbitrary gauge contains terms such as $T_Y(ux)T_Y(vx)$ ($Y\in\{\alpha,L,N\})$, and $T_N(x)$ oscillates at the sound speed (see Appendix for details). Here, we list some reasons why $I_{\chi}$ should not be physically taken as GWs.
\begin{itemize}
    \item Mathematically, $T_Y(ux)T_Y(vx)$ does not generate sinusoidal solutions such as $\sin x$ or $\cos x$, indicating the absence of wave-like propagation at the speed of light.
    \item In contrast to $I_X^{N}(u,v,x)$, which is formulated using the Green's function, $I_\chi(u,v,x)$ only contains the transfer function. The absence of the Green’s function implies that it lacks a propagation effect and thus does not represent physical waves.
    \item Analogous to quantum field theory, terms like $T_Y(ux)T_Y(vx)$ correspond to an off-shell contribution from the virtual modes rather than a final, on-shell observable graviton.
\end{itemize}
As a result, $I_\chi(u,v,x)$ consists of pure unphysical radiation regardless of the choice of $w$, $\alpha$ and $L$, indicating that $I_\chi(u,v,x)$ definitely violates the Sommerfeld boundary condition in the sub-horizon limit and should be discarded. 
Thus, the \emph{physical} $\Omega_{\mathrm{GW}}$ calculated in any gauge is always contributed by the first two oscillating components in Eq.~(\ref{Iuvsplit}) and yields the same result as that in the Newtonian gauge.

\begin{figure*}[htbp!]
\centering
\includegraphics[width=0.48\textwidth]{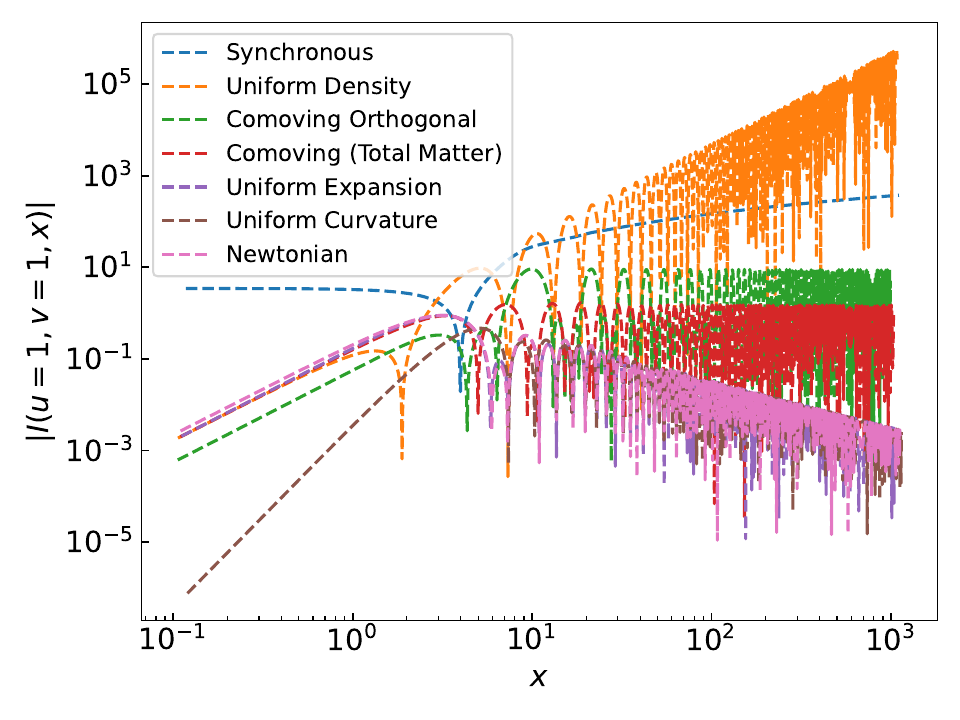}
\includegraphics[width =0.48\textwidth]{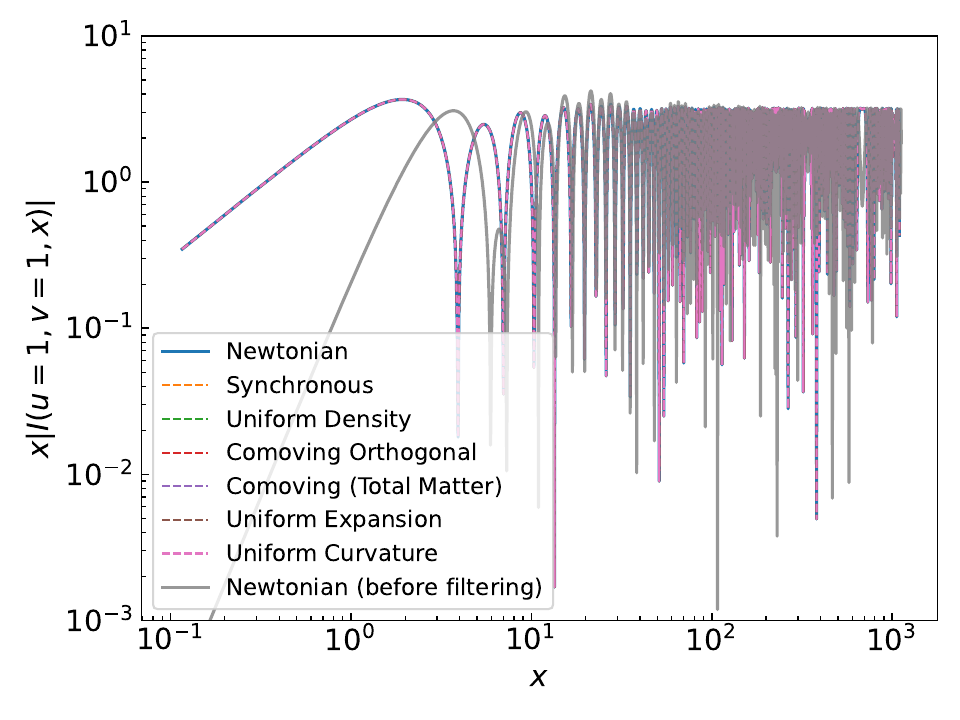}
\caption{\label{AD} Kernel functions of the second-order tensor modes for adiabatic perturbations in different gauges during RD.
\textit{Left panel}: The kernel functions of $h_{ij}$ without the boundary condition-based filtering.
\textit{Right panel}: The kernel functions of $h_{ij}$ after the filtering.
}
\end{figure*}

\begin{figure*}[htbp!]
\centering
\includegraphics[width=0.48\textwidth]{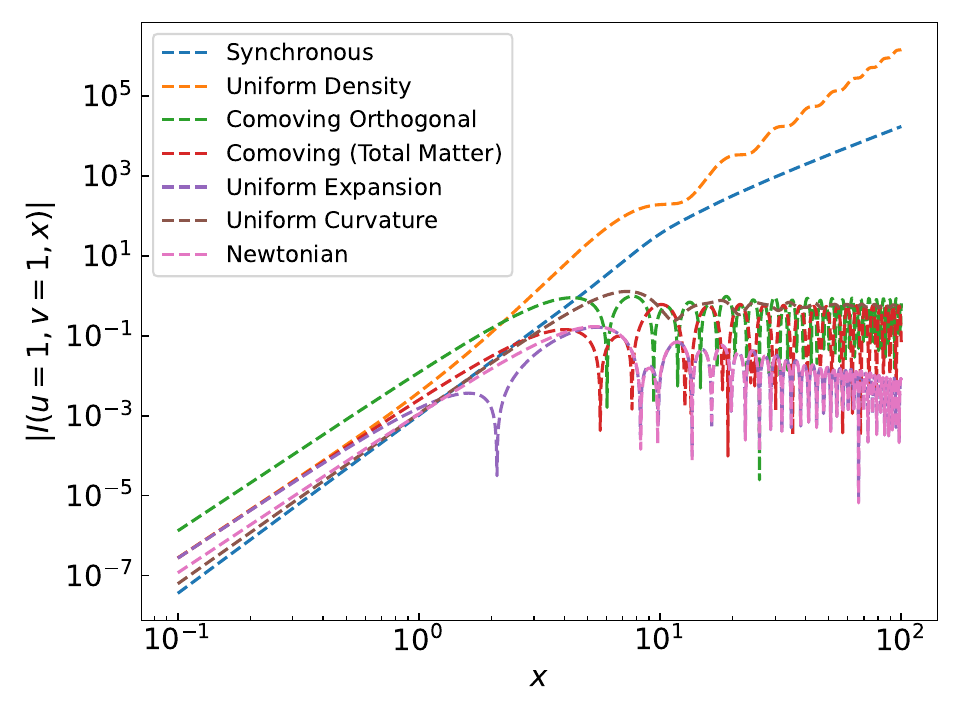}
\includegraphics[width =0.48\textwidth]{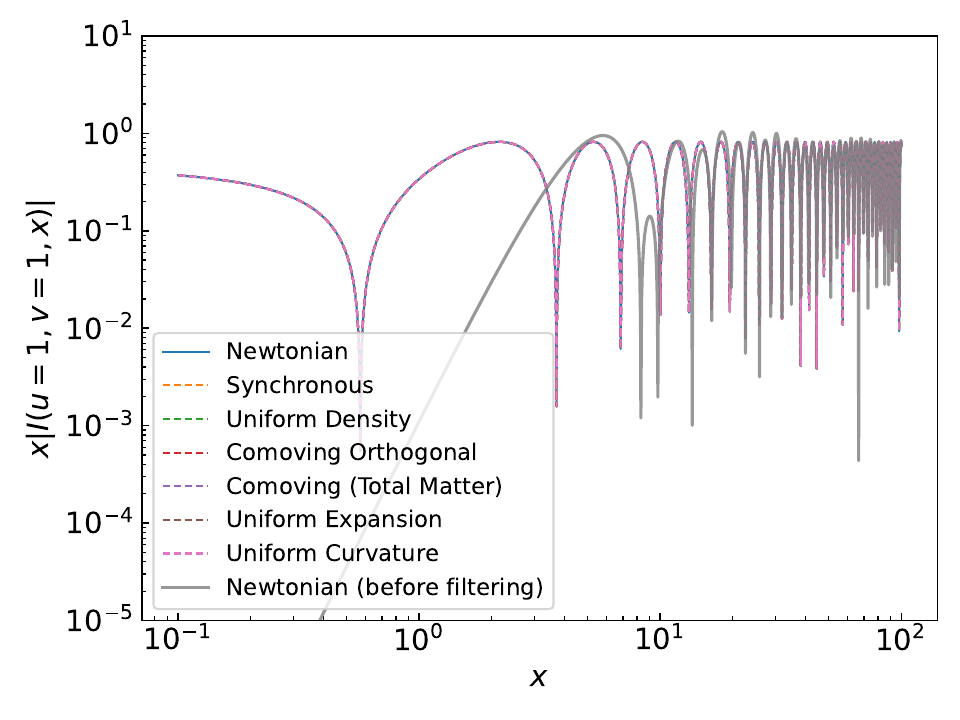}
\caption{\label{ISO} Kernel functions of the second-order tensor modes for isocurvature perturbations in different gauges during RD.
\textit{Left panel}: The kernel functions of $h_{ij}$ without the boundary condition-based filtering.
\textit{Right panel}: The kernel functions of $h_{ij}$ after the filtering.
}
\end{figure*}

Finally, we demonstrate the gauge invariance of physical $\Omega_{\rm GW}$ through concrete examples. For adiabatic perturbations, previous work~\cite{Lu:2020diy} showed that $\Omega_{\mathrm{GW}}$ in comoving gauge scales as $x^2$ in the sub-horizon limit. The transformation from Newtonian to comoving gauge is given by
\begin{equation}\label{comoving}
    T_\alpha(x) = -{1\over2}\left[xT_N(x)+x^2T_N’(x)\right]~,\quad T_L(x)=0.
\end{equation}
Using Eqs.~(\ref{comoving}) and (\ref{Ichi}), we obtain
\begin{equation}\label{comovingIchi}
\begin{aligned}
I_\chi=-&\frac{9}{8 u^3 v^3 x^4} \Bigg[ 
\Big( 6 u x \cos \frac{u x}{\sqrt{3}} 
+ \sqrt{3} \left( -6 + u^2 x^2 \right) \sin\frac{u x}{\sqrt{3}}  \Big) \\
&\times 
\Big( 6 v x \cos\frac{v x}{\sqrt{3}} 
+ \sqrt{3} \left( -6 + v^2 x^2 \right) \sin\frac{v x}{\sqrt{3}}  \Big) 
\Bigg].
\end{aligned}
\end{equation}
In the sub-horizon limit, this result leads to $I_\chi\sim \sin (ux/\sqrt{3})\sin (vx/\sqrt{3})$. The absence of light-speed propagating terms identifies this as an unphysical mode that must be discarded; retaining it would lead to the unphysical result $\Omega_{\mathrm{GW}} \sim x^2$~\cite{Lu:2020diy}. While~\cite{Domenech:2020xin} argues that comoving gauge is unsuitable for SIGW calculations due to its active source term continuously inducing tensor perturbations, we find that Eq.~(\ref{comovingIchi}) violates the Sommerfeld boundary condition in the sub-horizon limit. Consequently, $I_\chi$ vanishes under boundary condition-based filtering, ensuring that $\Omega_{\mathrm{GW}}$ observed by the canonical observer yields identical results in both comoving and Newtonian gauges. This demonstrates that SIGW calculations can proceed in any gauge, provided only physical modes contribute to $\Omega_{\mathrm{GW}}$, effectively rendering all gauges ``suitable'' for SIGW computation.

For the isocurvature case, the kernel function scales as $x^4$ in synchronous gauge \cite{Yuan:2024qfz}. To get into synchronous gauge from Newtonian gauge, one can choose
\begin{equation}\label{syn}
\begin{aligned}
        T_\alpha(x) &= \frac{9}{x} \left[ \frac{\sin\left( x/\sqrt{3}\right)}{x/\sqrt{3}} - 1 \right],\\
    T_{L}(x) &= 9 \left[ \mathcal{C} + \mathrm{Ci}\left( \frac{x}{\sqrt{3}} \right) - \log\left( \frac{x}{\sqrt{3}}\right) - \frac{\sin\left( x/\sqrt{3} \right)}{x/\sqrt{3}} \right],
\end{aligned}
\end{equation}
where $\mathcal{C}$ is a constant, due to the remaining gauge freedom in synchronous gauge. However, it can be seen that the transfer functions of $\alpha$ and $L$ oscillate at sound speed $x/\sqrt{3}$ and cannot produce terms propagating at the speed of light through $T_Y(ux)T_Y(vx)$ ($Y\in\{\alpha,L,N\})$. Therefore, $I_\chi$ would be filtered out from a canonical observer, and the resulting $\Omega_{\mathrm{GW}}$ coincides with the result in Newtonian gauge in the sub-horizon limit.

Finally, we present in Fig.~\ref{AD} and Fig.~\ref{ISO} for the results of kernel functions in different gauges. 
{For readers' convenience, we also show the analytical results for reproducing the right panels of Fig.~\ref{AD} and Fig.~\ref{ISO} in Appendix.~\ref{beforeafter}}
We demonstrate the results before and after the filtering. The definition of these gauges can be found in \cite{Lu:2020diy}. It can be seen that after the filtering, the kernel functions behave as $I\sim 1/x$ during RD, indicating that we successfully isolate only the physical components that behave as gravitational radiation in the second-order tensor modes.

\section{Discussion and Conclusion}

In this paper, we address the persistent gauge dependence issue in SIGWs through a novel boundary condition-based filtering method. While calculations in Newtonian gauge are well-established, the gauge dependence of $\Omega_{\mathrm{GW}}$ and its divergence in certain gauges have remained problematic. Our approach, based on the Sommerfeld criterion, establishes a unified framework for extracting physical gravitational radiation.
A key insight of our work, {which was first pointed out in \cite{Inomata:2019yww}}, is that the second-order transverse-traceless tensor $h_{ij}$ cannot be directly interpreted as GWs. Its gauge dependence introduces non-physical modes that violate fundamental properties of gravitational radiation: light-speed propagation and spherical wave behavior in the sub-horizon limit. We introduce the concept of a canonical observer who detects only physical modes satisfying the Sommerfeld boundary condition, with non-physical components naturally filtered out.

We demonstrated that, even in Newtonian gauge, there exists non-physical radiation denoted by $\Delta I_{X}^N(u,v,x)$. However, it will vanish in the sub-horizon limit, rendering $h_{ij}$ only contains physical modes in the late time.
It is also important to highlight a commonly overlooked issue in the calculation of GW energy, $\rho\sim\left\langle h_{ij}'h_{ij}'\right\rangle$. Specifically, when computing $\Omega_{\mathrm{GW}}$, it is customary to use $h_{ij}'(k)=kh_{ij}(k)$, which inherently presumes that the $h_{ij}$ represent waves that propagate at the speed of light. However, in certain gauges, the second-order tensor perturbations may include components that do not propagate at the speed of light and thus using $h_{ij}'(k)=kh_{ij}(k)$ would be paradoxically. Our proposed boundary condition-based filtering method effectively removes these non-physical components, ensuring the validity in calculating the energy of GWs.

By performing an arbitrary gauge transformation from Newtonian gauge and apply the boundary condition-based filtering method, we show that the resulting $\Omega_\mathrm{GW}$ observed by a canonical observer is convergent and gauge-invariant. This result holds for both adiabatic and isocurvature perturbations, providing a comprehensive solution to the gauge dependence problem.

Our approach complements and extends previous work on gauge suitability \cite{Domenech:2020xin}, where suitable gauges were identified based on inactive source term. While Newtonian gauge emerged as particularly suitable in that analysis, it did not resolve gauge dependence in arbitrary gauges with active sources. Our boundary condition-based filtering method provides a more general solution, applicable in any gauge while maintaining physical consistency.

This work also connects with recent developments in GW energy definitions. The strong conservation laws for curved backgrounds~\cite{Katz:1996nr} and canonical observer tetrads~\cite{Cai:2021jbi} suggest a deeper relationship between our filtering approach and fundamental gravitational physics. We conjecture that the tensor modes observed through canonical observer tetrads must satisfy the Sommerfeld boundary conditions in the sub-horizon limit, effectively implementing our filtering procedure. A rigorous proof of this correspondence presents an interesting direction for future research.

\section*{Acknowledgments}
This work has been supported by the National Key Research and Development Program of China (Grant No.~2023YFC2206704).
We are grateful to Xuan Ye and Xuefeng Feng for valuable discussions.
CY acknowledges financial support provided under the European Union’s H2020 ERC Advanced Grant “Black holes: gravitational engines of discovery” grant agreement no. Gravitas–101052587. 
Views and opinions expressed are however those of the author only and do not necessarily reflect those of the European Union or the European Research Council. Neither the European Union nor the granting authority can be held responsible for them.
This project has received funding from the European Union's Horizon 2020 research and innovation programme under the Marie Sklodowska-Curie grant agreement No 101007855 and No 101131233.
YL thanks SIMIS for support.
ZCC is supported by the National Natural Science Foundation of China under Grant No.~12405056, the Natural Science Foundation of Hunan Province under Grant No.~2025JJ40006, and the Innovative Research Group of Hunan Province under Grant No.~2024JJ1006. 
LL is supported by the National Natural Science Foundation of China Grant under Grant No. 12433001. 

\appendix

\section{Perturbed FLRW Metric up to Second Order}
The most general form of a perturbed FLRW metric up to second order can be written as
\begin{equation}
\begin{aligned}
d s^{2}&=a^{2}\big[-(1+2 \phi) \mathrm{d} \eta^{2} + 2(B_i+B_{,i}) \mathrm{d} x^i \mathrm{d} \eta \\
&+\left((1-2 \psi) \delta_{i j} + (E_{i,j}+E_{j,i})+2E_{,ij}
+ 2 h_{i j}^{(1)}
+\frac{1}{2} h_{i j}^{(2)}\right) \mathrm{d} x^{i} \mathrm{~d} x^{j}\big],
\end{aligned}
\end{equation}
where $\phi$, $\psi$, $B$, and $E$ denote the linear scalar perturbations, while $B_i$, $E_i$ represent the linear vector modes, and $h_{ij}^{(1)}$, $h_{ij}^{(2)}$ are the first- and second-order tensor modes, respectively. In our analysis, we neglect the linear vector and tensor modes.
We consider the radiation-dominated (RD) era, where the total stress-energy tensor takes the form $T_{\mu\nu}=T_{m\mu\nu}+T_{r\mu\nu}$. Here, $T_{m\mu\nu}$ and $T_{r\mu\nu}$ represent the stress-energy tensors of matter and radiation components, respectively, both described by perfect fluids. The purely adiabatic case corresponds to the limit $T_{m\mu\nu}\to 0$.

The metric perturbations are governed by the first-order Einstein equations:
\begin{equation}
\begin{aligned}\label{eom1}
    &\psi''+\mH\left[ \phi'+(2+3c_s^2)\psi')\right]
    +\left[ (1+3c_s^2)\mH^2 +2\mH' \right] \phi
    +c_s^2\left[\mH \Delta(B-E')-\Delta\psi\right]=4\pi a^2
\tau\delta s,\\
    &\phi-\psi+(B-E')'+2\mH(B-E')=0.
\end{aligned}
\end{equation}
Here, $\delta s$ is related to the pressure perturbation through $\delta p=c_s^2\delta \rho+\tau\delta s$ with the following definitions:
\begin{equation}
    c_s^2 = c_r^2\left(1+{\rho_m\over (1+w)\rho_r}\right)^{-1}, \qquad \tau={c_s^2 \rho_m \over s}, \qquad S\equiv {\delta s \over s }={\delta\rho_r\over (1+w)\rho_r}-{\delta\rho_m\over \rho_m}.
\end{equation}
Here $c_r$ is the speed of the radiation, and $S$ is the gauge invariant entropy perturbation. The transfer function of a general perturbation $X$ is defined by extracting the time evolution component in Fourier space from the perturbations and normalized as
\begin{equation}
    X=X_\mathbf{k}T_X(k\eta),
\end{equation}
where $X_\mathbf{k}$ is the primordial value of the perturbation with mode $\mathbf{k}$.

Consider a gauge transformation $x^\mu \to \tilde{x}^\mu =x^\mu +\xi^\mu$, where the gauge vector is decomposed as $\xi^\mu=(\alpha,\partial^{i}L)$. Under this transformation, the metric perturbations transform according to
\begin{equation}\label{transf}
    \begin{aligned}
    \tilde{\phi} & = \phi + \mH \alpha +\alpha',\\
    \tilde{\psi} & = \psi -\mH \alpha,\\
    \tilde{B} & = B- \alpha + L',\\
    \tilde{E} & = E + L.
    \end{aligned}
\end{equation}
Note that the Newtonian gauge ($\tilde{B}=\tilde{E}=0$) can be obtained by choosing $\alpha=B-E'$ and $L=-E$. An arbitrary gauge transformation from Newtonian gauge is given by
\begin{equation}\label{transf2}
    \begin{aligned}
    \tilde{\phi} & = \phi^N + \mH \alpha +\alpha',\\
    \tilde{\psi} & = \psi^N -\mH \alpha,\\
    \tilde{B} & = - \alpha + L',\\
    \tilde{E} & = L,\\
    \end{aligned}
\end{equation}
where $\phi^N=\psi^N$ represents the Bardeen potential in Newtonian gauge. For isocurvature case, $\phi^N$ is related to primordial entropy perturbation through $\phi^N_{\mathbf{k}}=S_{\mathbf{k}}$ while it is related to the primordial comoving curvature perturbation through \begin{equation}
    \phi^N_{\mathbf{k}}={\frac{3+3w}{5+3w}}\zeta_{\mathbf{k}}
\end{equation} for the adiabatic case on super-horizon scale. 

For adiabatic perturbations, $T_N(x)$ can be solved as
\begin{equation}
T_N(x) = 2^{\beta+1} \Gamma( \beta+2)(c_sx)^{-\beta-1}  J_{\beta+1}(c_sx) .
\end{equation}
For isocurvature perturbations in the very early universe, $T_N(x)$ is given by \cite{Domenech:2024wao}
\begin{align}
    T_N(x)=&  2^{-\frac{b}{2}-4} \left(1+b\right)^{b} \left(3-3 b^2\right)^{\frac{1}{4} (-2 b-3)} (k/k_{\mathrm{eq}}) ^{b-1} \Bigg[\frac{(1-b)^{b+\frac{5}{2}}}{b+4}x^{5/2} \Gamma \left(-b-\frac{3}{2}\right)  \nonumber \\ & \times 
    J_{-b-\frac{3}{2}}\left(c_w x\right) \, _1F_2\left(\frac{b}{2}+2;\frac{b}{2}+3,b+\frac{5}{2};-\frac{c_w^2 x^2}{4}\right)  \nonumber \\ &
    +12^{b+\frac{3}{2}} (b+1)^{b+\frac{3}{2}} x^{-2 b-\frac{1}{2}} \Gamma \left(b+\frac{3}{2}\right) \nonumber \\ & \times 
    J_{b+\frac{3}{2}}\left(c_w x\right) \, 
    _1F_2\left(\frac{1}{2}-\frac{b}{2};-b-\frac{1}{2},\frac{3}{2}-\frac{b}{2};-\frac{c_w^2 x^2}{4}\right)\Bigg]\,,
    \label{eq:Phi_Iso_wDE}
\end{align}
where $b=2\beta/3$ and $k_{\mathrm{eq}}$ is the comoving wavelength at equality.

Working in Newtonian gauge at second order, the tensor mode evolution is governed by Eq.~(1), where the quadratic source term takes the form
\begin{equation}
    S_{ij}=-\psi_{,i}\psi_{,j}-\phi_{,i}\phi_{,j}-2\psi_{,ij}(\phi+\psi)-
    \frac{2}{\mathcal{H'}-\mathcal{H}^2}\left(\mathcal{H}\phi+ \psi'\right)_{,i}
    \left(\mathcal{H}\phi+ \psi'\right)_{,j}.
\end{equation}
This source term maintains its form for both adiabatic and isocurvature perturbations. In Newtonian gauge, the scalar potentials are equal ($\psi=\phi$) when anisotropic stress is absent, further simplifying the expression.

In Fourier space, the source term can be written as
\begin{equation}
    S^\lambda(\eta,\mathbf{k})=4\int\frac{\mathrm{d}^3p}{(2\pi)^{3}} \Big(e^\lambda_{ij}p_i p_j\Big)X_\mathbf{p} X_{\mathbf{p}-\mathbf{k}} F(p,|\mathbf{k}-\mathbf{p}|,\eta),
\end{equation}
where the transfer function of the source term in Newtonian gauge is defined as
\begin{equation}
        F(u,v,x) = 2 T_N(ux)T_N(vx)+\frac{2}{\mH^2-\mH'}\left[\mH T_N(ux)+k \frac{dT_N(ux)}{dx} \right]
    \left[[\mH T_N(vx)+k \frac{dT_N(vx)}{dx}  \right].
\end{equation}

\section{{The kernel function of SIGWs during RD}}\label{beforeafter}
{During RD, $w=1/3$, the kernel function for adiabatic perturbations in Newtonian gauge before filtering is given by \cite{Kohri:2018awv}
\begin{align}\label{example-AD-N}
I_{\text{AD}}^{N}(v, u, x) =& \frac{3}{4 u^3 v^3 x} \left\{ -\frac{4}{x^3} \left[  uv (u^2 + v^2 -3) x^3 \sin x - 6 u v x^2 \cos \frac{ux}{\sqrt{3}} \cos\frac{vx}{\sqrt{3}} \right. \right. \nonumber \\
&\left.+ 6\sqrt{3} ux \cos \frac{ux}{\sqrt{3}} \sin \frac{vx}{\sqrt{3}} + 6 \sqrt{3} v x \sin \frac{ux}{\sqrt{3}}  \cos \frac{vx}{\sqrt{3}} \right. \nonumber \\
&\left. -3(6+(u^2+v^2-3)x^2) \sin\frac{ux}{\sqrt{3}}  \sin\frac{vx}{\sqrt{3}}   \right] \nonumber \\
& +(u^2+v^2-3)^2 \sin x \left[ \text{Ci}\left( \left( 1-\frac{v-u}{\sqrt{3}} \right)x \right) + \text{Ci}\left( \left( 1+\frac{v-u}{\sqrt{3}} \right)x \right) \right. \nonumber \\
& \left. - \text{Ci}\left( \left| 1-\frac{v+u}{\sqrt{3}} \right|x \right) - \text{Ci}\left( \left( 1+\frac{v+u}{\sqrt{3}} \right)x \right) + \log \left|\frac{3-(u+v)^2}{3-(u-v)^2} \right| \right] \nonumber \\
& +\cos x \left[ - \text{Si}\left( \left( 1-\frac{v-u}{\sqrt{3}} \right)x \right) -\text{Si}\left( \left( 1+\frac{v-u}{\sqrt{3}} \right)x \right)  \right. \nonumber \\
& \left. \left. + \mathrm{Si} \left( \left( 1-\frac{v+u}{\sqrt{3}} \right)x \right) + \mathrm{Si} \left( \left( 1+\frac{v+u}{\sqrt{3}} \right)x \right)   \right]  \right\}.
\end{align}
As we shown in Section.~\ref{method}, the non-physical modes do not satisfy the Sommerfeld boundary condition and need to be filtered out. This corresponds to the terms in the first three lines in \Eq{example-AD-N}, such as $\cos(ux/\sqrt{3})$, $\cos(vx/\sqrt{3})$, $\sin(ux/\sqrt{3})$ and $\sin(vx/\sqrt{3})$. All these terms do not propagate at the speed of light and do not behave as $1/x$ in the sub-horizon limit. Therefore we filter out these term in the subsequent computation and finally obtain
\begin{align}
I_{\text{AD}}(v, u, x) =& \frac{3}{4 u^3 v^3 x} \Bigg\{ -4   uv (u^2 + v^2 -3)  \sin x +(u^2+v^2-3)^2 \nonumber\\
&\times \sin x \left[ \text{Ci}\left( \left( 1-\frac{v-u}{\sqrt{3}} \right)x \right) +\text{Ci}\left( \left( 1+\frac{v-u}{\sqrt{3}} \right)x \right) \right. \nonumber \\
& \left. -\text{Ci}\left( \left| 1-\frac{v+u}{\sqrt{3}} \right|x \right) -\text{Ci}\left( \left( 1+\frac{v+u}{\sqrt{3}} \right)x \right) + \log \left|\frac{3-(u+v)^2}{3-(u-v)^2} \right| \right] \nonumber \\
& +\cos x \left[ -\text{Si}\left( \left( 1-\frac{v-u}{\sqrt{3}} \right)x \right) -\text{Si}\left( \left( 1+\frac{v-u}{\sqrt{3}} \right)x \right)  \right. \nonumber \\
& \left. + \text{Si}\left( \left( 1-\frac{v+u}{\sqrt{3}} \right)x \right) +\text{Si}\left( \left( 1+\frac{v+u}{\sqrt{3}} \right)x \right)   \right] \Bigg\},
\end{align}
which is also identical to the result of any gauges after filtering.}

{For isocurvature perturbations, it is difficult to obatin analytical result for $I_{\text{ISO}}^{N}(u,v,x)$, we calculate the kernel function numerically, see e.g., \cite{Yuan:2024qfz}. However, after filtering out the non-physical modes, one can analytically express the kernel function as 
\begin{align}\label{ISOafter}
I_{\text{ISO}}(u,v,x) &= \frac{9}{64 u^3 v^3 x \kappa^2} \Bigg[ -3u^2 v^2 + (-3 + u^2)(-3 + u^2 + 2v^2) \log\left| 1 - \frac{u^2}{3} \right| \nonumber \\
&\quad + (-3 + v^2)(-3 + v^2 + 2u^2) \log\left| 1 - \frac{v^2}{3} \right| \nonumber \\
&\quad - \frac{1}{2} (-3 + v^2 + u^2)^2 \log\left| \left( 1 - \frac{(u + v)^2}{3} \right)\left( 1 - \frac{(u - v)^2}{3} \right) \right| \Bigg] \cos{x}, \\
&+ \frac{9\pi}{32 u^3 v^3 x \kappa^2} \Bigg\{ 9 - 6v^2 - 6u^2 + 2u^2 v^2 + (3 - u^2)(-3 + u^2 + 2v^2)\, \Theta\left(1 - \frac{u}{\sqrt{3}}\right) \nonumber \\
&\quad + (3 - v^2)(-3 + v^2 + 2u^2)\, \Theta\left(1 - \frac{v}{\sqrt{3}}\right) \nonumber \\
&\quad + \frac{1}{2} (-3 + v^2 + u^2)^2 \left[ \Theta\left(1 - \frac{u + v}{\sqrt{3}}\right) + \Theta\left(1 + \frac{u - v}{\sqrt{3}}\right) \right] \Bigg\} \sin{x},
\end{align}
where $\kappa \equiv k/k_{\mathrm{eq}}\gg 1$ during RD, representing how deep are the perturbations inside the horizon. Eq.~(\ref{ISOafter}) is also identical to the result of any gauges after filtering.}

\bibliographystyle{JHEP}
\bibliography{ref}

\providecommand{\href}[2]{#2}\begingroup\raggedright\begin{thebibliography}{10}

\bibitem{LIGOScientific:2016dsl}
{\scshape LIGO Scientific, Virgo} collaboration, \emph{{Binary Black Hole Mergers in the first Advanced LIGO Observing Run}}, \href{https://doi.org/10.1103/PhysRevX.6.041015}{\emph{Phys. Rev. X} {\bfseries 6} (2016) 041015} [\href{https://arxiv.org/abs/1606.04856}{{\ttfamily 1606.04856}}].

\bibitem{LIGOScientific:2018mvr}
{\scshape LIGO Scientific, Virgo} collaboration, \emph{{GWTC-1: A Gravitational-Wave Transient Catalog of Compact Binary Mergers Observed by LIGO and Virgo during the First and Second Observing Runs}}, \href{https://doi.org/10.1103/PhysRevX.9.031040}{\emph{Phys. Rev. X} {\bfseries 9} (2019) 031040} [\href{https://arxiv.org/abs/1811.12907}{{\ttfamily 1811.12907}}].

\bibitem{LIGOScientific:2020ibl}
{\scshape LIGO Scientific, Virgo} collaboration, \emph{{GWTC-2: Compact Binary Coalescences Observed by LIGO and Virgo During the First Half of the Third Observing Run}}, \href{https://doi.org/10.1103/PhysRevX.11.021053}{\emph{Phys. Rev. X} {\bfseries 11} (2021) 021053} [\href{https://arxiv.org/abs/2010.14527}{{\ttfamily 2010.14527}}].

\bibitem{LIGOScientific:2021usb}
{\scshape LIGO Scientific, VIRGO} collaboration, \emph{{GWTC-2.1: Deep extended catalog of compact binary coalescences observed by LIGO and Virgo during the first half of the third observing run}}, \href{https://doi.org/10.1103/PhysRevD.109.022001}{\emph{Phys. Rev. D} {\bfseries 109} (2024) 022001} [\href{https://arxiv.org/abs/2108.01045}{{\ttfamily 2108.01045}}].

\bibitem{KAGRA:2021vkt}
{\scshape KAGRA, VIRGO, LIGO Scientific} collaboration, \emph{{GWTC-3: Compact Binary Coalescences Observed by LIGO and Virgo during the Second Part of the Third Observing Run}}, \href{https://doi.org/10.1103/PhysRevX.13.041039}{\emph{Phys. Rev. X} {\bfseries 13} (2023) 041039} [\href{https://arxiv.org/abs/2111.03606}{{\ttfamily 2111.03606}}].

\bibitem{Saito:2008jc}
R.~Saito and J.~Yokoyama, \emph{{Gravitational wave background as a probe of the primordial black hole abundance}}, \href{https://doi.org/10.1103/PhysRevLett.102.161101}{\emph{Phys. Rev. Lett.} {\bfseries 102} (2009) 161101} [\href{https://arxiv.org/abs/0812.4339}{{\ttfamily 0812.4339}}].

\bibitem{Sasaki:2018dmp}
M.~Sasaki, T.~Suyama, T.~Tanaka and S.~Yokoyama, \emph{{Primordial black holes\textemdash{}perspectives in gravitational wave astronomy}}, \href{https://doi.org/10.1088/1361-6382/aaa7b4}{\emph{Class. Quant. Grav.} {\bfseries 35} (2018) 063001} [\href{https://arxiv.org/abs/1801.05235}{{\ttfamily 1801.05235}}].

\bibitem{Chen:2019xse}
Z.-C.~Chen, C.~Yuan and Q.-G.~Huang, \emph{{Pulsar Timing Array Constraints on Primordial Black Holes with NANOGrav 11-Year Dataset}}, \href{https://doi.org/10.1103/PhysRevLett.124.251101}{\emph{Phys. Rev. Lett.} {\bfseries 124} (2020) 251101} [\href{https://arxiv.org/abs/1910.12239}{{\ttfamily 1910.12239}}].

\bibitem{Vaskonen:2020lbd}
V.~Vaskonen and H.~Veerm\"ae, \emph{{Did NANOGrav see a signal from primordial black hole formation?}}, \href{https://doi.org/10.1103/PhysRevLett.126.051303}{\emph{Phys. Rev. Lett.} {\bfseries 126} (2021) 051303} [\href{https://arxiv.org/abs/2009.07832}{{\ttfamily 2009.07832}}].

\bibitem{DeLuca:2020agl}
V.~De~Luca, G.~Franciolini and A.~Riotto, \emph{{NANOGrav Data Hints at Primordial Black Holes as Dark Matter}}, \href{https://doi.org/10.1103/PhysRevLett.126.041303}{\emph{Phys. Rev. Lett.} {\bfseries 126} (2021) 041303} [\href{https://arxiv.org/abs/2009.08268}{{\ttfamily 2009.08268}}].

\bibitem{Yuan:2021qgz}
C.~Yuan and Q.-G.~Huang, \emph{{A topic review on probing primordial black hole dark matter with scalar induced gravitational waves}}, \href{https://doi.org/10.1016/j.isci.2021.102860}{\emph{iScience} {\bfseries 24} (2021) 102860} [\href{https://arxiv.org/abs/2103.04739}{{\ttfamily 2103.04739}}].

\bibitem{Domenech:2021ztg}
G.~Dom\`enech, \emph{{Scalar Induced Gravitational Waves Review}}, \href{https://doi.org/10.3390/universe7110398}{\emph{Universe} {\bfseries 7} (2021) 398} [\href{https://arxiv.org/abs/2109.01398}{{\ttfamily 2109.01398}}].

\bibitem{Domenech:2023jve}
G.~Dom\`enech, \emph{{Cosmological gravitational waves from isocurvature fluctuations}}, \href{https://doi.org/10.1007/s43673-023-00109-z}{\emph{AAPPS Bull.} {\bfseries 34} (2024) 4} [\href{https://arxiv.org/abs/2311.02065}{{\ttfamily 2311.02065}}].

\bibitem{Matarrese:1997ay}
S.~Matarrese, S.~Mollerach and M.~Bruni, \emph{{Second order perturbations of the Einstein-de Sitter universe}}, \href{https://doi.org/10.1103/PhysRevD.58.043504}{\emph{Phys. Rev. D} {\bfseries 58} (1998) 043504} [\href{https://arxiv.org/abs/astro-ph/9707278}{{\ttfamily astro-ph/9707278}}].

\bibitem{Noh:2003yg}
H.~Noh and J.-c.~Hwang, \emph{{Second-order perturbations of the friedmann world model}},  \href{https://arxiv.org/abs/astro-ph/0305123}{{\ttfamily astro-ph/0305123}}.

\bibitem{Spiers:2023mor}
A.~Spiers, A.~Pound and B.~Wardell, \emph{{Second-order perturbations of the Schwarzschild spacetime: Practical, covariant, and gauge-invariant formalisms}}, \href{https://doi.org/10.1103/PhysRevD.110.064030}{\emph{Phys. Rev. D} {\bfseries 110} (2024) 064030} [\href{https://arxiv.org/abs/2306.17847}{{\ttfamily 2306.17847}}].

\bibitem{Gong:2019mui}
J.-O.~Gong, \emph{{Analytic Integral Solutions for Induced Gravitational Waves}}, \href{https://doi.org/10.3847/1538-4357/ac3a6c}{\emph{Astrophys. J.} {\bfseries 925} (2022) 102} [\href{https://arxiv.org/abs/1909.12708}{{\ttfamily 1909.12708}}].

\bibitem{Tomikawa:2019tvi}
K.~Tomikawa and T.~Kobayashi, \emph{{Gauge dependence of gravitational waves generated at second order from scalar perturbations}}, \href{https://doi.org/10.1103/PhysRevD.101.083529}{\emph{Phys. Rev. D} {\bfseries 101} (2020) 083529} [\href{https://arxiv.org/abs/1910.01880}{{\ttfamily 1910.01880}}].

\bibitem{DeLuca:2019ufz}
V.~De~Luca, G.~Franciolini, A.~Kehagias and A.~Riotto, \emph{{On the Gauge Invariance of Cosmological Gravitational Waves}}, \href{https://doi.org/10.1088/1475-7516/2020/03/014}{\emph{JCAP} {\bfseries 03} (2020) 014} [\href{https://arxiv.org/abs/1911.09689}{{\ttfamily 1911.09689}}].

\bibitem{Inomata:2019yww}
K.~Inomata and T.~Terada, \emph{{Gauge Independence of Induced Gravitational Waves}}, \href{https://doi.org/10.1103/PhysRevD.101.023523}{\emph{Phys. Rev. D} {\bfseries 101} (2020) 023523} [\href{https://arxiv.org/abs/1912.00785}{{\ttfamily 1912.00785}}].

\bibitem{Yuan:2019fwv}
C.~Yuan, Z.-C.~Chen and Q.-G.~Huang, \emph{{Scalar induced gravitational waves in different gauges}}, \href{https://doi.org/10.1103/PhysRevD.101.063018}{\emph{Phys. Rev. D} {\bfseries 101} (2020) 063018} [\href{https://arxiv.org/abs/1912.00885}{{\ttfamily 1912.00885}}].

\bibitem{Lu:2020diy}
Y.~Lu, A.~Ali, Y.~Gong, J.~Lin and F.~Zhang, \emph{{Gauge transformation of scalar induced gravitational waves}}, \href{https://doi.org/10.1103/PhysRevD.102.083503(2020)}{\emph{Phys. Rev. D} {\bfseries 102} (2020) 083503} [\href{https://arxiv.org/abs/2006.03450}{{\ttfamily 2006.03450}}].

\bibitem{Ali:2020sfw}
A.~Ali, Y.~Gong and Y.~Lu, \emph{{Gauge transformation of scalar induced tensor perturbation during matter domination}}, \href{https://doi.org/10.1103/PhysRevD.103.043516}{\emph{Phys. Rev. D} {\bfseries 103} (2021) 043516} [\href{https://arxiv.org/abs/2009.11081}{{\ttfamily 2009.11081}}].

\bibitem{Yuan:2024qfz}
C.~Yuan, Z.-C.~Chen and L.~Liu, \emph{{Gauge Dependence of Gravitational Waves Induced by Primordial Isocurvature Fluctuations}},  \href{https://arxiv.org/abs/2410.18996}{{\ttfamily 2410.18996}}.

\bibitem{Hwang:2017oxa}
J.-C.~Hwang, D.~Jeong and H.~Noh, \emph{{Gauge dependence of gravitational waves generated from scalar perturbations}}, \href{https://doi.org/10.3847/1538-4357/aa74be}{\emph{Astrophys. J.} {\bfseries 842} (2017) 46} [\href{https://arxiv.org/abs/1704.03500}{{\ttfamily 1704.03500}}].

\bibitem{Domenech:2020xin}
G.~Dom\`enech and M.~Sasaki, \emph{{Approximate gauge independence of the induced gravitational wave spectrum}}, \href{https://doi.org/10.1103/PhysRevD.103.063531}{\emph{Phys. Rev. D} {\bfseries 103} (2021) 063531} [\href{https://arxiv.org/abs/2012.14016}{{\ttfamily 2012.14016}}].

\bibitem{Cai:2021jbi}
R.-G.~Cai, X.-Y.~Yang and L.~Zhao, \emph{{On the energy of gravitational waves}}, \href{https://doi.org/10.1007/s10714-022-02972-x}{\emph{Gen. Rel. Grav.} {\bfseries 54} (2022) 89} [\href{https://arxiv.org/abs/2109.06864}{{\ttfamily 2109.06864}}].

\bibitem{Newman:1961qr}
E.~Newman and R.~Penrose, \emph{{An Approach to gravitational radiation by a method of spin coefficients}}, \href{https://doi.org/10.1063/1.1724257}{\emph{J. Math. Phys.} {\bfseries 3} (1962) 566}.

\bibitem{Kohri:2018awv}
K.~Kohri and T.~Terada, \emph{{Semianalytic calculation of gravitational wave spectrum nonlinearly induced from primordial curvature perturbations}}, \href{https://doi.org/10.1103/PhysRevD.97.123532}{\emph{Phys. Rev. D} {\bfseries 97} (2018) 123532} [\href{https://arxiv.org/abs/1804.08577}{{\ttfamily 1804.08577}}].

\bibitem{Domenech:2019quo}
G.~Dom\`enech, \emph{{Induced gravitational waves in a general cosmological background}}, \href{https://doi.org/10.1142/S0218271820500285}{\emph{Int. J. Mod. Phys. D} {\bfseries 29} (2020) 2050028} [\href{https://arxiv.org/abs/1912.05583}{{\ttfamily 1912.05583}}].

\bibitem{Domenech:2024wao}
G.~Dom\`enech and J.~Tr\"ankle, \emph{{From formation to evaporation: Induced gravitational wave probes of the primordial black hole reheating scenario}},  \href{https://arxiv.org/abs/2409.12125}{{\ttfamily 2409.12125}}.

\bibitem{Domenech:2021and}
G.~Dom\`enech, S.~Passaglia and S.~Renaux-Petel, \emph{{Gravitational waves from dark matter isocurvature}}, \href{https://doi.org/10.1088/1475-7516/2022/03/023}{\emph{JCAP} {\bfseries 03} (2022) 023} [\href{https://arxiv.org/abs/2112.10163}{{\ttfamily 2112.10163}}].

\bibitem{Maggiore:2007ulw}
M.~Maggiore, \emph{{Gravitational Waves. Vol. 1: Theory and Experiments}}, Oxford University Press (2007), \href{https://doi.org/10.1093/acprof:oso/9780198570745.001.0001}{10.1093/acprof:oso/9780198570745.001.0001}.

\bibitem{Katz:1996nr}
J.~Katz, J.~Bicak and D.~Lynden-Bell, \emph{{Relativistic conservation laws and integral constraints for large cosmological perturbations}}, \href{https://doi.org/10.1103/PhysRevD.55.5957}{\emph{Phys. Rev. D} {\bfseries 55} (1997) 5957} [\href{https://arxiv.org/abs/gr-qc/0504041}{{\ttfamily gr-qc/0504041}}].

\end{thebibliography}\endgroup
\end{document}